%% file: pglpm190808-evidence_vs_scores.tex
\pdfoutput=1
\newif\ifarxiv
\arxivtrue
\ifarxiv\pdfmapfile{+classico.map}\fi
\newif\ifafour
\afourfalse
\newif\iftypodisclaim 
\typodisclaimtrue

\documentclass[\ifafour a4paper,12pt,\else a5paper,10pt,\fi
onecolumn,oneside,article,
british%
]{memoir}

\newcommand*{\firstpublished}{18 August 2019}
\newcommand*{\updated}{20 August 2019}
\newcommand*{\propertitle}{A relation between log-likelihood\\ and cross-validation
  log-scores
}
\newcommand*{\pdftitle}{A relation between log-likelihood and cross-validation log-scores}
\newcommand*{\headtitle}{log-likelihood and cross-validation}
\newcommand*{\pdfauthor}{P.G.L.  Porta Mana}
\newcommand*{\headauthor}{Porta Mana}
\newcommand*{\reporthead}{\ifarxiv\else Open Science Framework \href{https://doi.org/10.31219/osf.io/k8mj3}{\textsc{doi}:10.31219/osf.io/k8mj3}\fi}

\usepackage[T1]{fontenc} 
\input{glyphtounicode} \pdfgentounicode=1

\usepackage[utf8]{inputenx}

\usepackage{textcomp}

\usepackage{amsmath}

\usepackage{mathtools}
\usepackage{empheq}

\usepackage{amssymb}

\usepackage{amsxtra}

\usepackage[main=british,french,italian,german,swedish,latin,esperanto]{babel}

\usepackage[autostyle=false,autopunct=false,english=british]{csquotes}
\setquotestyle{british}

\usepackage{amsthm}

\theoremstyle{remark}

\newtheoremstyle{innote}{\parsep}{\parsep}{\footnotesize}{}{}{}{0pt}{}
\theoremstyle{innote}
\newtheorem*{innote}{}

\usepackage[shortlabels,inline]{enumitem}
\SetEnumitemKey{para}{itemindent=\parindent,leftmargin=0pt,listparindent=\parindent,parsep=0pt,itemsep=\topsep}
\setlist[enumerate,2]{label=\alph*.}
\setlist[enumerate]{label=\arabic*.,leftmargin=1.5\parindent}
\setlist[itemize]{leftmargin=1.5\parindent}
\setlist[description]{leftmargin=1.5\parindent}

\usepackage[babel,theoremfont,largesc]{newpxtext}

\usepackage[bigdelims,nosymbolsc
]{newpxmath}
\makeatletter
\def\re@DeclareMathSymbol#1#2#3#4{%
    \let#1=\undefined
    \DeclareMathSymbol{#1}{#2}{#3}{#4}}
\re@DeclareMathSymbol{\bigoplusop}{\mathop}{largesymbols}{"4C}
\re@DeclareMathSymbol{\bigotimesop}{\mathop}{largesymbols}{"4E}
\re@DeclareMathSymbol{\sumop}{\mathop}{largesymbols}{"50}
\re@DeclareMathSymbol{\prodop}{\mathop}{largesymbols}{"51}
\re@DeclareMathSymbol{\bigcupop}{\mathop}{largesymbols}{"53}
\re@DeclareMathSymbol{\bigcapop}{\mathop}{largesymbols}{"54}
\re@DeclareMathSymbol{\bigwedgeop}{\mathop}{largesymbols}{"56}
\re@DeclareMathSymbol{\bigveeop}{\mathop}{largesymbols}{"57}
\re@DeclareMathSymbol{\bigtimesop}{\mathop}{largesymbolsPXA}{"10}
\makeatother
\DeclareFontFamily{U}{egreek}{\skewchar\font'177}%
\DeclareFontShape{U}{egreek}{m}{n}{<-6>s*[1]eurm5 <6-8>s*[1]eurm7 <8->s*[1]eurm10}{}%
\DeclareFontShape{U}{egreek}{m}{it}{<->s*[1]eurmo10}{}%
\DeclareFontShape{U}{egreek}{b}{n}{<-6>s*[1]eurb5 <6-8>s*[1]eurb7 <8->s*[1]eurb10}{}%
\DeclareFontShape{U}{egreek}{b}{it}{<->s*[1]eurbo10}{}%
\DeclareSymbolFont{egreeki}{U}{egreek}{m}{it}%
\SetSymbolFont{egreeki}{bold}{U}{egreek}{b}{it}
\DeclareSymbolFont{egreekr}{U}{egreek}{m}{n}%
\SetSymbolFont{egreekr}{bold}{U}{egreek}{b}{n}
\DeclareFontFamily{U}{egreekx}{\skewchar\font'177}
\DeclareFontShape{U}{egreekx}{m}{n}{%
       <-7.5>s*[0.9]euex7%
    <7.5-8.5>s*[0.9]euex8%
    <8.5-9.5>s*[0.9]euex9%
    <9.5->s*[0.9]euex10%
}{}
\DeclareSymbolFont{egreekx}{U}{egreekx}{m}{n}
\DeclareMathSymbol{\sumop}{\mathop}{egreekx}{"50}
\DeclareMathSymbol{\prodop}{\mathop}{egreekx}{"51}
\DeclareMathSymbol{\coprodop}{\mathop}{egreekx}{"60}
\makeatletter
\def\sum{\DOTSI\sumop\slimits@}
\def\prod{\DOTSI\prodop\slimits@}
\def\coprod{\DOTSI\coprodop\slimits@}
\makeatother
\input{definegreek.tex}

\renewcommand\sfdefault{uop}
\DeclareMathAlphabet{\mathsf}  {T1}{\sfdefault}{m}{sl}
\SetMathAlphabet{\mathsf}{bold}{T1}{\sfdefault}{b}{sl}

\usepackage[scaled=0.84]{DejaVuSansMono}

\usepackage{mathdots}

\usepackage[usenames]{xcolor}
\definecolor{mypurpleblue}{RGB}{68,119,170}
\definecolor{myblue}{RGB}{102,204,238}
\definecolor{mygreen}{RGB}{34,136,51}
\definecolor{myyellow}{RGB}{204,187,68}
\definecolor{myred}{RGB}{238,102,119}
\definecolor{myredpurple}{RGB}{170,51,119}
\definecolor{mygrey}{RGB}{187,187,187}
\definecolor{lgrey}{RGB}{221,221,221}
\colorlet{shadecolor}{lgrey}

\usepackage{bm}

\usepackage{microtype}

\usepackage[backend=biber,mcite,
citestyle=authoryear-comp,bibstyle=pglpm-authoryear,autopunct=false,sorting=ny,sortcites=false,natbib=false,maxcitenames=1,maxbibnames=8,minbibnames=8,giveninits=true,uniquename=false,uniquelist=false,maxalphanames=1,block=space,hyperref=true,defernumbers=false,useprefix=true,sortupper=false,language=british,parentracker=false]{biblatex}
\DeclareSortingScheme{ny}{\sort{\field{sortname}\field{author}\field{editor}}\sort{\field{year}}}
\DefineBibliographyExtras{british}{\def\finalandcomma{\addcomma}}
\renewcommand*{\finalnamedelim}{\addcomma\space}
\DeclareDelimFormat{multicitedelim}{\addcomma\space}
\DeclareDelimFormat{compcitedelim}{\addcomma\space}
\DeclareDelimFormat{postnotedelim}{\space}
\ifarxiv\else\fi
\renewcommand{\bibfont}{\footnotesize}
\defbibheading{bibliography}[\bibname]{\section*{#1}\addcontentsline{toc}{section}{#1}
}
\newcommand*{\citep}{\footcites}
\newcommand*{\citey}{\footcites}
\providecommand{\href}[2]{#2}

\newcommand*{\amp}{\&}

\newcommand*{\subtitleproc}[1]{}

\usepackage{graphicx}

\PassOptionsToPackage{hyphens}{url}\usepackage[hypertexnames=false]{hyperref}

\usepackage[depth=4]{bookmark}
\hypersetup{colorlinks=true,bookmarksnumbered,pdfborder={0 0 0.25},citebordercolor={0.2667 0.4667 0.6667},citecolor=mypurpleblue,linkbordercolor={0.6667 0.2 0.4667},linkcolor=myredpurple,urlbordercolor={0.1333 0.5333 0.2},urlcolor=mygreen,breaklinks=true,pdftitle={\pdftitle},pdfauthor={\pdfauthor}}

\usepackage[british]{datetime2}
\DTMnewdatestyle{mydate}%
{
}
\DTMsetdatestyle{mydate}

\ifafour\setstocksize{297mm}{210mm}
\else\setstocksize{210mm}{5.5in}
\fi
\settrimmedsize{\stockheight}{\stockwidth}{*}
\setlxvchars[\normalfont] 
\setxlvchars[\normalfont]
\ifafour\settypeblocksize{*}{32pc}{1.618} 
\else\settypeblocksize{*}{26pc}{1.618}
\fi
\setulmargins{*}{*}{1}
\setlrmargins{*}{*}{*}
\setheadfoot{\onelineskip}{2.5\onelineskip}
\setheaderspaces{*}{2\onelineskip}{*}
\setmarginnotes{2ex}{10mm}{0pt}
\checkandfixthelayout[nearest]
\fixpdflayout
\pdfmapline{+dummy-space <dummy-space.pfb}\pdfinterwordspaceon%

\newenvironment{acknowledgements}{\section*{Thanks}\addcontentsline{toc}{section}{Thanks}}{\par}
\makeatletter\renewcommand{\appendix}{\par
  \bigskip{\centering
   \interlinepenalty \@M
   \normalfont
   \printchaptertitle{\sffamily\appendixpagename}\par}
  \setcounter{section}{0}%
  \gdef\@chapapp{\appendixname}%
  \gdef\thesection{\@Alph\c@section}%
  \anappendixtrue}\makeatother
\counterwithout{section}{chapter}
\setsecnumformat{\upshape\csname the#1\endcsname\quad}
\setsecheadstyle{\large\bfseries\sffamily%
\centering}
\setsubsecheadstyle{\bfseries\sffamily%
\raggedright}
\setsubsecindent{0pt}
\setparaheadstyle{\bfseries\sffamily%
\raggedright}
\newcommand{\addchap}[1]{\chapter*[#1]{#1}\addcontentsline{toc}{chapter}{#1}}
\newcommand{\addsec}[1]{\section*{#1}\addcontentsline{toc}{section}{#1}}

\copypagestyle{manaart}{plain}
\makeheadrule{manaart}{\headwidth}{0.5\normalrulethickness}
\makeoddhead{manaart}{%
{\footnotesize
\scshape\headauthor}}{}{{\footnotesize\sffamily%
\headtitle}}
\makeoddfoot{manaart}{}{\thepage}{}
\newcommand*\autanet{\includegraphics[height=\heightof{M}]{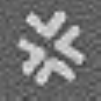}}
\definecolor{mygray}{gray}{0.333}
\iftypodisclaim%
\ifafour\newcommand\addprintnote{\begin{picture}(0,0)%
\put(245,149){\makebox(0,0){\rotatebox{90}{\tiny\color{mygray}\textsf{This
            document is designed for screen reading and
            two-up printing on A4 or Letter paper}}}}%
\end{picture}}
\else\newcommand\addprintnote{\begin{picture}(0,0)%
\put(176,112){\makebox(0,0){\rotatebox{90}{\tiny\color{mygray}\textsf{This
            document is designed for screen reading and
            two-up printing on A4 or Letter paper}}}}%
\end{picture}}\fi
\makeoddfoot{plain}{}{\makebox[0pt]{\thepage}\addprintnote}{}
\else
\makeoddfoot{plain}{}{\makebox[0pt]{\thepage}}{}
\fi
\makeoddhead{plain}{\scriptsize\reporthead}{}{}

\pretitle{\begin{center}\LARGE\sffamily%
\bfseries}
\posttitle{\bigskip\end{center}}

\makeatletter\newcommand*{\atf}{\includegraphics[
totalheight=\heightof{@}]{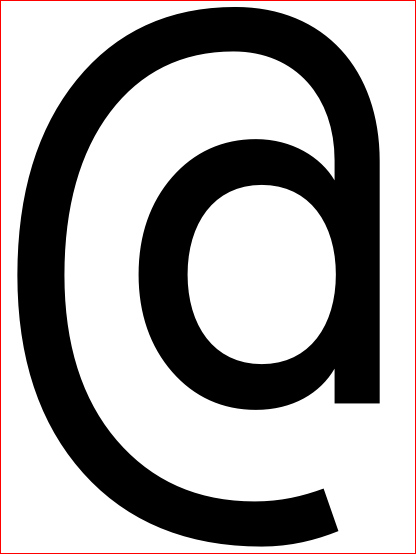}}\makeatother

\providecommand{\epost}[1]{\texttt{\footnotesize\textless#1\textgreater}}
\providecommand{\email}[2]{\href{mailto:#1ZZ@#2 ((remove ZZ))}{#1\protect\atf#2}}

\preauthor{\vspace{-0.5\baselineskip}\begin{center}
\normalsize\sffamily%
\lineskip  0.5em}
\postauthor{\par\end{center}}
\predate{\DTMsetdatestyle{mydate}\begin{center}\footnotesize}
\postdate{\end{center}\vspace{-\medskipamount}}

\setfloatadjustment{figure}{\footnotesize}
\captiondelim{\quad}
\captionnamefont{\footnotesize\sffamily%
}
\captiontitlefont{\footnotesize}
\firmlists*
\midsloppy
\paragraphfootnotes
\footmarkstyle{\textsuperscript{\color{myred}\bfseries#1}~}

\title{\propertitle}
\author{%
\hspace*{\stretch{1}}%
\parbox{0.75\linewidth}
{\protect\centering P.G.L.  Porta Mana\\%
\footnotesize Kavli Institute, Trondheim, Norway \quad\epost{\email{piero.mana}{ntnu.no}}}%
\hspace*{\stretch{1}}%
}

\date{\firstpublished; updated \updated}

\newcommand*{\I}{\mathrm{i}}
\newcommand*{\e}{\mathrm{e}}

\newcommand*{\defd}{\coloneqq}

\newcommand*{\Land}{\bigwedge}

\DeclarePairedDelimiter\set{\{}{\}}
\newcommand*{\p}{\mathrm{P}}
\newcommand*{\E}{\mathrm{E}}
\renewcommand*{\|}[1][]{\nonscript\,#1\vert\nonscript\;\mathopen{}}
\DeclarePairedDelimiterX{\condt}[2]{[}{]}{#1\nonscript\,\delimsize\vert\nonscript\;\mathopen{}#2}
\newcommand*{\sect}{\S}
\newcommand*{\sects}{\S\S}
\newcommand*{\chap}{ch.}%
\newcommand*{\chaps}{chs}%
\newcommand*{\eqn}{eq.}%

\newcommand*{\etc}{{etc.}}
\newcommand*{\eg}{{e.g.}}

\newcommand*{\tlor}{\mathop{\textstyle\bigvee}\nolimits}

 \definecolor{notecolour}{RGB}{68,170,153}

\newcommand*{\dob}{degree of belief}

\newcommand*{\yK}{I}
\newcommand*{\yO}{\mathrm{O}}

\firmlists
\begin{document}
\captiondelim{\quad}\captionnamefont{\footnotesize}\captiontitlefont{\footnotesize}
\selectlanguage{british}\frenchspacing
\maketitle

\abstractrunin
\abslabeldelim{}
\renewcommand*{\abstractname}{}
\setlength{\absleftindent}{0pt}
\setlength{\absrightindent}{0pt}
\setlength{\abstitleskip}{-\absparindent}
\begin{abstract}\labelsep 0pt%
  \noindent It is shown that the log-likelihood of a hypothesis or model
  given some data is equal to an average of all leave-one-out
  cross-validation log-scores that can be calculated from all subsets of
  the data. This relation can be generalized to any $k$-fold
  cross-validation log-scores.
\\\noindent\emph{\footnotesize Note: Dear Reader
    \amp\ Peer, this manuscript is being peer-reviewed by you. Thank you.}
\end{abstract}
\selectlanguage{british}\frenchspacing



\section{Log-likelihoods and cross-validation log-scores}
\label{sec:intro}

The probability calculus unequivocally tells us how our \dob\ in a
hypothesis $H_{h}$ given data $D$ and background information or assumptions
$\yK$, that is, $\p(H_{h} \| D \, \yK)$, is related to our \dob\ in
observing those data when we entertain that hypothesis as true, that is,
$\p(D \| H_{h} \, \yK)$:
\begin{subequations}
    \label{eq:posterior_hypothesis}
  \begin{align}
    \label{eq:posterior_hypothesis_universal}
    \p(H_{h} \| D \, \yK) &=
    \frac{\p(D \| H_{h} \, \yK)\;\p(H_{h} \| \yK)}{\p(D \| \yK)}\\
    \label{eq:posterior_hypothesis_sethypotheses}
    &=\frac{\p(D \| H_{h} \, \yK)\;\p(H_{h} \| \yK)}{\sum_{h'} \p(D \| H_{h'} \, \yK)\; \p(H_{h'} \| \yK)}.
  \end{align}
\end{subequations}
$D$, $H_{h}$, $\yK$ denote propositions, which are usually about
numeric quantities. I use the terms \enquote{\dob}, \enquote{belief}, and
\enquote{probability} as synonyms. By \enquote{hypothesis} I mean either a
scientific (physical, biological, \etc) hypothesis -- a state or
development of things capable of experimental verification, at least in a
thought experiment -- or more generally some proposition, often not
precisely specified, which leads to quantitatively specific distributions
of beliefs for any contemplated data set.
In the latter case we often call $H_{h}$  a
\enquote{(probabilistic) model} rather than a \enquote{hypothesis}.
  

Expression~\eqref{eq:posterior_hypothesis_sethypotheses} assumes that we
have a set $\set{H_{h}}$ of mutually exclusive and exhaustive hypotheses
under consideration, which is implicit in our knowledge $\yK$. In fact it's
only valid if
\begin{equation}
  \label{eq:implicit_knowledge}
  \p\bigl(\tlor_{h} H_{h} \| \yK\bigr) = 1,
  \qquad
  \p(H_{h} \land H_{h'} \| \yK) = 0 \quad \text{if $h \ne h'$}.
\end{equation}\pagebreak
Only  rarely does the set of hypotheses $\set{H_{h}}$
encompass and reflect the extremely complex and fuzzy hypotheses lying in
the backs of our minds. They're simplified pictures. That's also why
they're 
called \enquote{models}.

Expression~\eqref{eq:posterior_hypothesis_universal} is universally valid
instead, but it's rarely possible to quantify its denominator
$\p(D \| \yK)$ unless we simplify our inferential problem by introducing a
possibly unrealistic exhaustive set of hypotheses, thus falling back
to~\eqref{eq:posterior_hypothesis_sethypotheses}. We can bypass this
problem if we are content with comparing our beliefs about any two
hypotheses through their ratio, so that the term $\p(D \| \yK)$ cancels
out. See Jaynes's \citey[\sects~4.3--4.4]{jaynes1994_r2003} insightful
remarks about such binary comparisons, and also Good's
\citey[\sect~6.3--6.6]{good1950}.

\bigskip

The term $\p(D \| H_{h} \, \yK)$ in \eqn~\eqref{eq:posterior_hypothesis} is
called the \emph{likelihood} of the hypothesis given the data
\citep[\sect~6.1 p.~62]{good1950}. Its logarithm is surprisingly called
log-likelihood:
\begin{equation}
  \label{eq:log-likelihood}
  \log\p(D \| H_{h} \, \yK),
\end{equation}
where the logarithm can be taken in an arbitrary basis (Turing, Good
\citep[\eg][]{good1985,good1950,good1969}, Jaynes
\citep[\sect~4.2]{jaynes1994_r2003} recommend base
$\cramped10^{1/10}$, leading to a measurement in decibels; see the
cited works for the practical advantages of such choice).

The ratio of the likelihoods of two hypotheses, called \emph{relative Bayes
  factor}, or its logarithm, the \emph{relative weight of
  evidence},\footnote{\cites[\chap~6]{good1950}{good1975,good1981,good1985},
  and many other works in \cite{good1983};
  \cites[\sect~1.4]{osteyeeetal1974}{mackay1992,kassetal1995}; see also
  \cite[\chaps~V, VI, A]{jeffreys1939_r1983}.} are often used to quantify
how much the data favour our belief in one versus the other
hypothesis (that is, assuming at least momentarily that they be
exhaustive). \enquote{It is historically interesting that the expression
  ``weight of evidence'', in its technical sense, anticipated the term
  ``likelihood'' by over forty years} \citep[\sect~1.4.2
p.~12]{osteyeeetal1974}.
\begin{innote}
  Recent literature \citep[for example][]{kassetal1995} seems to
  exclusively deal with \emph{relative} Bayes factors. I'd like to recall,
  lest it fades from the memory, the definition of the non-relative Bayes
  factor for a hypothesis $H_{h}$ provided by data
  $D$:\citep[\sect~2]{good1981}
  \begin{equation}
    \label{eq:proper_Bayes_factor}
    \frac{\p(D \| H_{h} \; \yK)}{\p(D \| \lnot H_{h} \; \yK)} \equiv
    \frac{\yO(H_{h} \| D \; \yK)}{\yO(H_{h} \| \yK)} =
    \frac{\p(D \| H_{h}\; \yK)\; [1- \p(H_{h} \| \yK)]}{
\sum_{h'}^{h' \ne h} \p(D \| H_{h'} \; \yK) \; \p(H_{h'} \| \yK)
    },
  \end{equation}
  where the \emph{odds} $\yO$ is defined as $\yO \defd \p/(1-\p)$. Looking
  at the expression on the right, which can be derived from the probability
  rules, it's clear that the Bayes factor for a hypothesis involves the
  likelihoods of \emph{all} other hypotheses as well as their pre-data
  probabilities. This quantity and its logarithm, the (non-relative) weight
  of evidence, have important properties which relative Bayes factors and
  relative weights of evidence don't enjoy. For example, the expected
  weight of evidence for a correct hypothesis is always positive, and for a
  wrong hypotheses always negative\citep[\sect~6.7]{good1950}. See
  Jaynes\citey[\sects~4.3--4.4]{jaynes1994_r2003} for further discussion
  and a numeric example.
\end{innote}

\bigskip

The literature in probability and statistics has also employed and debated
other ad-hoc measures to quantify how the data relate to the hypotheses --
or even to select one hypothesis for further use, discarding the
others\citep[\sects~3.4, 6.1.6 gives the clearest motivation and
explanation]{bernardoetal1994_r2000}[see
also][]{stone1977,geisseretal1979,vehtarietal2012,vehtarietal2002,krnjajicetal2011,krnjajicetal2014,gelmanetal2014,gronauetal2019,chandramoulietal2019}.
Here I consider one measure in particular: the \emph{leave-one-out
  cross-validation log-score}\addtocounter{footnote}{-1}\footnotemark{},
which I'll just call \enquote{log-score} for brevity:
\begin{equation}
  \label{eq:log-score}
  \frac{1}{d} \sum_{i=1}^{d} \log\p(D_{i} \| D_{-i} \, H_{h} \, \yK)
\end{equation}
where every $D_{i}$ is one datum in the data $D \equiv \Land_{i=1}^{d} D_{i}$,
and $D_{-i}$ denotes the data with datum $D_{i}$ excluded. The intuition
behind this score can be colloquially expressed thus: \enquote{let's see
  what my belief in one datum would be, on average, once I've observed the
  other data, if I consider $H_{h}$ as true}. \enquote{On average} means
considering such belief for every single datum in turn, and then taking the
geometric mean of the resulting
beliefs. 
Other variants of this score use more general partitions of the data into
two disjoint subsets\addtocounter{footnote}{-1}\footnotemark{}.

\textcolor{white}{If you find this you can claim a postcard from me.}

My purpose is to show an exact relation between the
log-likelihood~\eqref{eq:log-likelihood} and the leave-one-out
cross-validation log-score~\eqref{eq:log-score}. This relation doesn't seem
to appear in the literature, and I find it very intriguing because it
portrays the log-likelihood as a sort of full-scale use of the log-score:
it says that \emph{the log-likelihood is the sum of all averaged log-scores
  that can be formed from all data subsets}. The relation can be extended
to more general cross-validation log-scores, and it can be of interest for
the debate about the soundness of log-scores in deciding among hypotheses.

\section{A relation between log-likelihood and log-score}
\label{sec:relation}




\bigskip

We can obviously write the likelihood as the $d$th root of its $d$th power:
\begin{equation}
  \label{eq:root_product}
  \p(D \| H \, \yK) \equiv  \bigl[\,
  \underbracket{\p(D \| H \, \yK) \times \dotsm \times
  \p(D \| H \, \yK)}_{\text{$d$ times}}
  \,\bigr]^{1/d}
\end{equation}
where we have dropped the subscript ${}_{h}$ for simplicity. By the rules
of probability we have
\begin{equation}
  \label{eq:product_rule}
  \p(D \| H \, \yK) =
  \p(D_{i} \| D_{-i} \, H_{h} \, \yK) \times \p(D_{-i} \|  H_{h} \, \yK)
\end{equation}
no matter which specific $i \in \set{1, \dotsc, d}$ we choose (temporal
ordering and similar matters are completely irrelevant in the formula
above: it's a logical relation between propositions). So let's expand each
of the $d$ factors in the identity~\eqref{eq:root_product} using the
product rule~\eqref{eq:product_rule}, using a different $i$ for each of
them. The result can be thus displayed:
\begin{equation}
  \label{eq:product_2}
  \begin{aligned}
    \p(D \| H \, \yK) \equiv{}
    \bigl[\,&\p(D_{1} \| D_{-1} \, H \, \yK) \times
            \p(D_{-1} \|  H \, \yK) \times{}\\
          &\p(D_{2} \| D_{-2} \, H \, \yK) \times
            \p(D_{-2} \| H \, \yK)\times{}\\
          &\hphantom{\p(D_{2} \| D_{-2} \, H \, \yK)}
            \mathrlap{\dotso}\hphantom{{}\times  \p(D_{-2} \| H \, \yK)}\times{}\\
          &\p(D_{d} \|\underbracket[0pt][1ex]{ D_{-d} }_{\mathclap{%
              \substack{\Big\uparrow\\\text{this column leads to the log-score}}%
}} H \, \yK) \times\p(D_{-d} \|  H \, \yK)
            \,\bigr]^{1/d}.
  \end{aligned}
\end{equation}
Upon taking the logarithm of this expression, the $d$ factors vertically
aligned on the left add up to the log-score~\eqref{eq:log-score}, as
indicated. But the mathematical reshaping we just did for
$\p(D \| H \, \yK)$ -- that is, the root-product
identity~\eqref{eq:root_product} and the expansion~\eqref{eq:product_2} --
can be done for each of the remaining factors $\p(D_{-i} \| H \, \yK)$
vertically aligned on the right in the expression above; and so on
recursively. Here is an explicit example for $d=3$:
\begin{multline}
  \label{eq:example_further_expansion}
  \p(D \| H \, \yK) \equiv{}\\
  \begin{alignedat}[b]{2}
    \Bigl\{\,&\p(D_{1} \| D_{2} \,D_{3} \, H \, \yK) \times
    \bigl[ &&\p(D_{2} \| D_{3} \, H \, \yK) \times \p( D_{3} \| H \, \yK) \times{}\\[-0.5\jot]
    &&&\p(D_{3} \| D_{2} \, H \, \yK) \times \p( D_{2} \| H \, \yK)
    \bigr]^{1/2}\times{}\\[\jot]
    &\p(D_{2} \| D_{1} \,D_{3} \, H \, \yK) \times
    \bigl[&&\p(D_{1} \| D_{3} \, H \, \yK) \times \p( D_{3} \| H \, \yK) \times{}\\[-0.5\jot]
    &&&\p(D_{3} \| D_{1} \, H \, \yK) \times \p( D_{1} \| H \, \yK)
    \bigr]^{1/2}\times{}\\[\jot]
    &\p(D_{3} \| D_{1} \,D_{2} \, H \, \yK) \times
    \bigl[&&\p(D_{1} \| D_{2} \, H \, \yK) \times \p( D_{2} \| H \, \yK) \times{}\\[-0.5\jot]
    &&&\p(D_{2} \| D_{1} \, H \, \yK) \times \p( D_{1} \| H \, \yK)
    \bigr]^{1/2}\Bigr\}^{1/3}.
  \end{alignedat}
\end{multline}
In this example the logarithm of the three vertically aligned factors in
the left column is, as already noted, the log-score~\eqref{eq:log-score}.
The logarithm of the six vertically aligned factors in the central column
is an average of the log-scores calculated for the three distinct subsets
of pairs of data $\set{D_{1}\,D_{2}}$, $\set{D_{1}\, D_{3}}$,
$\set{D_{2}\, D_{3}}$. Likewise, the logarithm of the six factors
vertically aligned on the right is the average of the log-scores for the
three subsets of data singletons $\set{D_{1}}$, $\set{D_{2}}$,
$\set{D_{3}}$.

In the general case with $d$ data there are $\binom{d}{k}$ subsets with $k$
data points. We therefore obtain
\begin{multline}
  \label{eq:general_equivalence}
  \log\p(D \| H \, \yK) \equiv
  \frac{1}{d} \sum_{i=1}^{d} \log\p(D_{i} \| D_{-i} \, H \, \yK)
+{}\\
\shoveright{\frac{1}{d} \sum_{i\in\set{1,\dotsc,d}} \frac{1}{d-1} \sum_{j\in\set{1,\dotsc,d}}^{j\ne i}  \!\!\! \log\p(D_{-i,j} \| D_{-i,-j} \, H \, \yK)
  +{}}\\[0.5\jot]
\shoveright{\binom{d}{d-2}^{-1} \!\!\! \sum_{i,j\in\set{1,\dotsc,d}}^{i<j} \frac{1}{d-2} \sum_{k\in\set{1,\dotsc,d}}^{k\ne i,j} \!\!\! \log\p(D_{-i,-j,k} \| D_{-i,-j,-k} \, H \, \yK)
      +{}}\\[\jot]
       {} \dotsb +{}\\
\shoveright{      \binom{d}{2}^{-1} \!\!\! \sum_{i,j\in\set{1,\dotsc,d}}^{i<j} \frac{1}{2}
      \bigl[ \log\p(D_{i} \| D_{j} \, H \, \yK) +
      \log\p(D_{j} \| D_{i} \, H \, \yK)\bigr]
  +{}}\\[0.5\jot]
  \frac{1}{d} \sum_{i=1}^{d} \log\p(D_{i} \| H \, \yK),
\end{multline}
which can be compactly written
\begin{empheq}[box=\fbox]{equation}
    \label{eq:general_equivalence_compact}
    \log\p(D \| H \, \yK) \equiv
    \sum_{k=1}^{d}
    \binom{d}{k}^{-1}
    \!\smashoperator{\sum_{\substack{\text{ordered}\\\text{$k$-tuples}}}} 
    \;\;\;\frac{1}{k}\;\;
    \smashoperator{\sum_{\substack{\text{cyclic}\\\text{permutations}}}} 
    \log\p(D_{i_{1}} \| D_{i_{2}} \dotsm D_{i_{k}} \, H \, \yK).
\end{empheq}
That is, \emph{the log-likelihood is the sum of all averaged log-scores
  that can be formed from all (non-empty) data subsets with $k$ elements},
the average for log-scores over $k$ data being taken over the
$\binom{d}{k}$ subsets having the same cardinality $k$.

There's also an equivalent form with a slightly different cross-validating
interpretation: We take each datum $D_{j}$ in turn and calculate our
log-belief in it conditional on all possible subsets of remaining data,
from the empty subset with no data (term $k=0$), to the only subset
$D_{-j}$ with all data except $D_{j}$ (term $k=d-1$). These log-beliefs are
averaged over the $\binom{d-1}{k}$ subsets having the same cardinality $k$.
The result can be expressed as
\begin{empheq}[box=\fbox]{equation}
    \label{eq:general_equivalence_alternative}
    \log\p(D \| H \, \yK) \equiv
    \frac{1}{d}\sum_{j=1}^{d}
    \sum_{k=0}^{d-1}
    \binom{d-1}{k}^{-1}
    \smashoperator{\sum_{\substack{\text{ordered}\\\text{$k$-tuples,}\\\text{$j$ excluded}}}}
    \log\p(D_{j} \| D_{i_{1}} \dotsm D_{i_{k}} \, H \, \yK).
\end{empheq}

\section{Brief discussion}
\label{sec:discussion}


It's remarkable that the individual log-scores in
expressions~\eqref{eq:general_equivalence_compact} and
\eqref{eq:general_equivalence_alternative} above are computationally
expensive, but their sum results in the
log-likelihood, which is less expensive.

The relation~\eqref{eq:general_equivalence_compact} invites us to see the
log-likelihood as a refinement and improvement of the log-score. The
log-likelihood takes into account not only the log-score for the whole
data, but also the log-scores for all possible subsets of data.
Figuratively speaking it examines the relationship between data and
hypothesis locally, globally, and on all intermediate scales. To me this
property makes the log-likelihood preferable to any single log-score
(besides the fact that the log-likelihood is directly obtained from the
principles of the probability calculus), because our interest is usually in how
the hypothesis $H$ relates to single data points as well as to any
collection of them. I hope to discuss this point, which also involves the
distinction between simple and composite
hypotheses\citep[\sect~6.1.4]{bernardoetal1994_r2000}, more in detail
elsewhere\citep{portamana9999b}.

By applying the identity~\eqref{eq:root_product} and generalizing the
expansion~\eqref{eq:product_rule} to different divisions of the data --
leave-two-out, leave-three-out, and so on -- we see that the
relation~\eqref{eq:general_equivalence_compact} can be generalized to any
$k$-fold cross-validation log-scores. Thus the log-likelihood is also
equivalent to an average of \emph{all conceivable} cross-validation
log-scores for all subsets of data, though I haven't calculated the weights
of such average.

\begin{acknowledgements}
  \ldots to the Kavli Foundation and the Centre of Excellence scheme of the
  Research Council of Norway (Yasser Roudi group) for financial support. To
  Aki Vehtari for some references. To the staff of the NTNU library for
  their always prompt assistance. To Mari, Miri, Emma for continuous
  encouragement and affection, and to Buster Keaton and Saitama for filling
  life with awe and inspiration. To the developers and maintainers of Open
  Science Framework, \LaTeX, Emacs, AUC\TeX, Python, Inkscape, Sci-Hub for
  making a free and impartial scientific exchange possible.%
\\\mbox{}\hfill\autanet
\end{acknowledgements}


\defbibnote{prenote}{{\footnotesize (\enquote{de $X$} is listed under D,
    \enquote{van $X$} under V, and so on, regardless of national
    conventions.)\par}}

\printbibliography[
]

\end{document}

%% file: definegreek.tex
\DeclareMathSymbol{\varpartial}{\mathalpha}{egreeki}{"40}
\DeclareMathSymbol{\partialup}{\mathalpha}{egreekr}{"40}
\DeclareMathSymbol{\alpha}{\mathalpha}{egreeki}{"0B}
\DeclareMathSymbol{\beta}{\mathalpha}{egreeki}{"0C}
\DeclareMathSymbol{\gamma}{\mathalpha}{egreeki}{"0D}
\DeclareMathSymbol{\delta}{\mathalpha}{egreeki}{"0E}
\DeclareMathSymbol{\epsilon}{\mathalpha}{egreeki}{"0F}
\DeclareMathSymbol{\zeta}{\mathalpha}{egreeki}{"10}
\DeclareMathSymbol{\eta}{\mathalpha}{egreeki}{"11}
\DeclareMathSymbol{\theta}{\mathalpha}{egreeki}{"12}
\DeclareMathSymbol{\iota}{\mathalpha}{egreeki}{"13}
\DeclareMathSymbol{\kappa}{\mathalpha}{egreeki}{"14}
\DeclareMathSymbol{\lambda}{\mathalpha}{egreeki}{"15}
\DeclareMathSymbol{\mu}{\mathalpha}{egreeki}{"16}
\DeclareMathSymbol{\nu}{\mathalpha}{egreeki}{"17}
\DeclareMathSymbol{\xi}{\mathalpha}{egreeki}{"18}
\DeclareMathSymbol{\omicron}{\mathalpha}{egreeki}{"6F}
\DeclareMathSymbol{\pi}{\mathalpha}{egreeki}{"19}
\DeclareMathSymbol{\rho}{\mathalpha}{egreeki}{"1A}
\DeclareMathSymbol{\sigma}{\mathalpha}{egreeki}{"1B}
 \DeclareMathSymbol{\tau}{\mathalpha}{egreeki}{"1C}
\DeclareMathSymbol{\upsilon}{\mathalpha}{egreeki}{"1D}
\DeclareMathSymbol{\phi}{\mathalpha}{egreeki}{"1E}
\DeclareMathSymbol{\chi}{\mathalpha}{egreeki}{"1F}
\DeclareMathSymbol{\psi}{\mathalpha}{egreeki}{"20}
\DeclareMathSymbol{\omega}{\mathalpha}{egreeki}{"21}
\DeclareMathSymbol{\varepsilon}{\mathalpha}{egreeki}{"22}
\DeclareMathSymbol{\vartheta}{\mathalpha}{egreeki}{"23}
\DeclareMathSymbol{\varpi}{\mathalpha}{egreeki}{"24}
\let\varrho\rho 
\let\varsigma\sigma
 \let\varkappa\kappa
\DeclareMathSymbol{\varphi}{\mathalpha}{egreeki}{"27}
\DeclareMathSymbol{\varAlpha}{\mathalpha}{egreeki}{"41}
\DeclareMathSymbol{\varBeta}{\mathalpha}{egreeki}{"42}
\DeclareMathSymbol{\varGamma}{\mathalpha}{egreeki}{"00}
\DeclareMathSymbol{\varDelta}{\mathalpha}{egreeki}{"01}
\DeclareMathSymbol{\varEpsilon}{\mathalpha}{egreeki}{"45}
\DeclareMathSymbol{\varZeta}{\mathalpha}{egreeki}{"5A}
\DeclareMathSymbol{\varEta}{\mathalpha}{egreeki}{"48}
\DeclareMathSymbol{\varTheta}{\mathalpha}{egreeki}{"02}
 \DeclareMathSymbol{\varIota}{\mathalpha}{egreeki}{"49}
\DeclareMathSymbol{\varKappa}{\mathalpha}{egreeki}{"4B}
\DeclareMathSymbol{\varLambda}{\mathalpha}{egreeki}{"03}
\DeclareMathSymbol{\varMu}{\mathalpha}{egreeki}{"4D}
\DeclareMathSymbol{\varNu}{\mathalpha}{egreeki}{"4E}
\DeclareMathSymbol{\varXi}{\mathalpha}{egreeki}{"04}
\DeclareMathSymbol{\varOmicron}{\mathalpha}{egreeki}{"4F}
\DeclareMathSymbol{\varPi}{\mathalpha}{egreeki}{"05}
\DeclareMathSymbol{\varRho}{\mathalpha}{egreeki}{"50}
\DeclareMathSymbol{\varSigma}{\mathalpha}{egreeki}{"06}
\DeclareMathSymbol{\varTau}{\mathalpha}{egreeki}{"54}
\DeclareMathSymbol{\varUpsilon}{\mathalpha}{egreeki}{"07}
\DeclareMathSymbol{\varPhi}{\mathalpha}{egreeki}{"08}
\DeclareMathSymbol{\varChi}{\mathalpha}{egreeki}{"58}
\DeclareMathSymbol{\varPsi}{\mathalpha}{egreeki}{"09}
\DeclareMathSymbol{\varOmega}{\mathalpha}{egreeki}{"0A} 
\DeclareMathSymbol{\Alpha}{\mathalpha}{egreekr}{"41}
\DeclareMathSymbol{\Beta}{\mathalpha}{egreekr}{"42}
\DeclareMathSymbol{\Gamma}{\mathalpha}{egreekr}{"00}
\DeclareMathSymbol{\Delta}{\mathalpha}{egreekr}{"01}
\DeclareMathSymbol{\Epsilon}{\mathalpha}{egreekr}{"45}
\DeclareMathSymbol{\Zeta}{\mathalpha}{egreekr}{"5A}
\DeclareMathSymbol{\Eta}{\mathalpha}{egreekr}{"48}
\DeclareMathSymbol{\Theta}{\mathalpha}{egreekr}{"02}
\DeclareMathSymbol{\Iota}{\mathalpha}{egreekr}{"49}
\DeclareMathSymbol{\Kappa}{\mathalpha}{egreekr}{"4B}
\DeclareMathSymbol{\Lambda}{\mathalpha}{egreekr}{"03}
\DeclareMathSymbol{\Mu}{\mathalpha}{egreekr}{"4D}
\DeclareMathSymbol{\Nu}{\mathalpha}{egreekr}{"4E}
\DeclareMathSymbol{\Xi}{\mathalpha}{egreekr}{"04}
\DeclareMathSymbol{\Omicron}{\mathalpha}{egreekr}{"4F}
\DeclareMathSymbol{\Pi}{\mathalpha}{egreekr}{"05}
\DeclareMathSymbol{\Rho}{\mathalpha}{egreekr}{"50}
\DeclareMathSymbol{\Sigma}{\mathalpha}{egreekr}{"06}
\DeclareMathSymbol{\Tau}{\mathalpha}{egreekr}{"54}
\DeclareMathSymbol{\Upsilon}{\mathalpha}{egreekr}{"07}
\DeclareMathSymbol{\Phi}{\mathalpha}{egreekr}{"08}
\DeclareMathSymbol{\Chi}{\mathalpha}{egreekr}{"58}
\DeclareMathSymbol{\Psi}{\mathalpha}{egreekr}{"09}
\DeclareMathSymbol{\Omega}{\mathalpha}{egreekr}{"0A}
\DeclareMathSymbol{\alphaup}{\mathalpha}{egreekr}{"0B}
\DeclareMathSymbol{\betaup}{\mathalpha}{egreekr}{"0C}
\DeclareMathSymbol{\gammaup}{\mathalpha}{egreekr}{"0D}
 \DeclareMathSymbol{\deltaup}{\mathalpha}{egreekr}{"0E}
\DeclareMathSymbol{\epsilonup}{\mathalpha}{egreekr}{"0F}
\DeclareMathSymbol{\zetaup}{\mathalpha}{egreekr}{"10}
\DeclareMathSymbol{\etaup}{\mathalpha}{egreekr}{"11}
\DeclareMathSymbol{\thetaup}{\mathalpha}{egreekr}{"12}
\DeclareMathSymbol{\iotaup}{\mathalpha}{egreekr}{"13}
\DeclareMathSymbol{\kappaup}{\mathalpha}{egreekr}{"14}
\DeclareMathSymbol{\lambdaup}{\mathalpha}{egreekr}{"15}
\DeclareMathSymbol{\muup}{\mathalpha}{egreekr}{"16}
\DeclareMathSymbol{\nuup}{\mathalpha}{egreekr}{"17}
\DeclareMathSymbol{\xiup}{\mathalpha}{egreekr}{"18}
\DeclareMathSymbol{\omicronup}{\mathalpha}{egreekr}{"6F}
  \DeclareMathSymbol{\piup}{\mathalpha}{egreekr}{"19}
\DeclareMathSymbol{\rhoup}{\mathalpha}{egreekr}{"1A}
\DeclareMathSymbol{\sigmaup}{\mathalpha}{egreekr}{"1B}
\DeclareMathSymbol{\tauup}{\mathalpha}{egreekr}{"1C}
\DeclareMathSymbol{\upsilonup}{\mathalpha}{egreekr}{"1D}
\DeclareMathSymbol{\phiup}{\mathalpha}{egreekr}{"1E}
\DeclareMathSymbol{\chiup}{\mathalpha}{egreekr}{"1F}
\DeclareMathSymbol{\psiup}{\mathalpha}{egreekr}{"20}
\DeclareMathSymbol{\omegaup}{\mathalpha}{egreekr}{"21}
\DeclareMathSymbol{\varepsilonup}{\mathalpha}{egreekr}{"22}
\DeclareMathSymbol{\varthetaup}{\mathalpha}{egreekr}{"23}
\DeclareMathSymbol{\varpiup}{\mathalpha}{egreekr}{"24}

\DeclareMathSymbol{\varphiup}{\mathalpha}{egreekr}{"27}